\documentclass[10pt, conference]{IEEEtran}

\makeatletter
\def\ps@headings{%
\def\@oddhead{\mbox{}\scriptsize\rightmark \hfil \thepage}%
\def\@evenhead{\scriptsize\thepage \hfil \leftmark\mbox{}}%
\def\@oddfoot{}
\def\@evenfoot{}}
\def\blfootnote{\xdef\@thefnmark{}\@footnotetext}
\makeatother
\usepackage{graphicx} 
\usepackage{epsfig} 
\usepackage{color}
\usepackage{mathptmx} 
\usepackage{times} 
\usepackage{latexsym}
\usepackage{amsmath} 
\usepackage{amssymb}  
\usepackage{subfigure}
\usepackage{algorithmic}
\usepackage{setspace}
\usepackage[bottom]{footmisc}
\newtheorem{theorem}{Theorem}

\newtheorem{lemma}{Lemma}
\newtheorem{definition}{Definition}

\title{\huge \bf
Optimal computation of symmetric Boolean functions in Tree networks
}

\author{\IEEEauthorblockN{Hemant Kowshik}
\IEEEauthorblockA{CSL and Department of ECE\\
University of Illinois Urbana-Champaign\\
Email: kowshik2@illinois.edu}
\and
\IEEEauthorblockN{P. R. Kumar}
\IEEEauthorblockA{CSL and Department of ECE\\
University of Illinois Urbana-Champaign\\
Email: prkumar@illinois.edu}
}
\begin{document}
\maketitle\blfootnote{This material is based upon work partially supported by AFOSR under Contract FA9550-09-0121, NSF under Contract Nos. CNS-05-19535, CNS-07-21992, ECCS-0701604,
and CNS-0626584, and USARO under Contract Nos. W911NF-08-1-0238 and W-911-NF-0710287,
Any opinions, findings, and conclusions or recommendations expressed in this publication are those of the authors and do not necessarily reflect the views of the above agencies.}
\thispagestyle{empty}
\pagestyle{empty}
\begin{abstract}
In this paper, we address the scenario where nodes with sensor data are connected in a tree network, and every node wants to compute a given symmetric Boolean function of the sensor data. We first consider the problem of computing a function of two nodes with integer measurements. We allow for block computation to enhance data fusion efficiency, and determine the minimum worst-case total number of bits to be exchanged to perform the desired computation. We establish lower bounds using fooling sets, and provide a novel scheme which attains the lower bounds, using information theoretic tools. For a class of functions called sum-threshold functions, this scheme is shown to be optimal. 

We then turn to tree networks and derive a lower bound for the number of bits exchanged on each link by viewing it as a two node problem. We show that the protocol of recursive in-network aggregation achieves this lower bound in the case of sum-threshold functions. Thus we have provided a communication and in-network computation strategy that is optimal for each link. All the results can be extended to the case of non-binary alphabets.  

In the case of general graphs, we present a cut-set lower bound, and an achievable scheme based on aggregation along trees. For complete graphs, the complexity of this scheme is no more than twice that of the optimal scheme.
\end{abstract}
\section{INTRODUCTION}
Wireless sensor networks are composed of nodes with sensing, wireless communication and computation capabilities. 
In sensor network applications, one is not interested in the raw sensor measurements, but only in computing some relevant \textit{function} of the measurements. For example, one might want to compute the mean temperature for environmental monitoring, or the maximum temperature in fire alarm systems. Thus, we need to move away from a data-centric paradigm, and focus on efficient in-network computation and communication strategies which are function-aware. 

There are two possible architectures for sensor networks that one might consider. First, one can designate a single collector node which seeks to compute the function of interest. Alternately, one can suppose that \textit{every} node in the network wants to compute the function. 
The latter architecture can be viewed as providing implicit feedback to each sensor node, which could be useful in applications like fault monitoring and adaptive sensing. For example, sensor nodes could modify their sampling rate depending on the value of the function. In this paper, we will focus on strategies which achieve function computation with zero error for \textit{all} sensor nodes. 

In this paper, we abstract out the medium access control problem, and view the network as a  graph with edges representing essentially noiseless wired links between nodes. Hence, we focus on strategies for combining information at intermediate nodes, and optimal codes for transmissions on each edge. This is a significant departure from the traditional decode and forward paradigm in wireless networks. Moreover, the strategy for computation may benefit from interactive information exchange between nodes.

We consider a graph where each node has a Boolean variable and focus on the problem of symmetric Boolean function computation. We adopt a deterministic formulation of the problem of function computation, allowing zero error. We consider the problem of worst-case function computation, without imposing a probability distribution on the node measurements. Further, instead of restricting a strategy to compute just one instance of the problem, we allow for nodes to accumulate a block of $B$ measurements, and realize greater                                                                                                                                                                                                                                                        efficiency by using block codes. The set of admissible strategies includes all interactive strategies, where a node may exchange several messages with other nodes, with node $i$'s transmission being allowed to depend on all previous transmissions heard by node $i$, and node $i$'s block of measurements.


We begin with the two node problem in Section \ref{sec_two_node}, where each node $i$ has an integer variable $X_i$ and both nodes want to compute a function $f(X_1, X_2)$ which only depends on $X_1 + X_2$. We use a lower bound from the theory of communication complexity, by constructing an appropriate fooling set \cite{KushiNisan}. For achievability, we devise a single-round strategy 
so as to minimize the worst-case total number of bits exchanged under the Kraft inequality constraint. 
For the class of sum-threshold functions, which evaluate to 1 if $X_1 + X_2$ exceeds a threshold, this single-round strategy is indeed optimal. However, for the class of sum-interval functions, which evaluate to 1 if $a \leq X_1 + X_2 \leq b$, the upper and lower bounds do not match. However, the achievable strategy involving separation, followed by coding, can be used for any general function.

In Section \ref{sec_trees}, we consider Boolean symmetric function computation on trees. Since every edge is a cut-edge, we can obtain a cut-set lower bound for the number of bits that must be exchanged on an edge, by reducing it to a two node problem. 
For the class of sum-threshold functions, we are able to match the cut-set bound by constructing an achievable strategy that is reminiscent of message passing algorithms.

In Section \ref{sec_gen_graphs}, for general graphs, we can still derive a cut-set lower bound by considering all partitions of the vertices. We also propose an achievable scheme that consists of activating a subtree of edges and using the optimal strategy for transmissions on the tree. While the upper and lower bounds do not match even for very simple functions, for complete graphs we show that aggregation along trees provides a 2-OPT solution.
\section{RELATED WORK}
The problem of function computation in sensor networks has received much attention recently. In \cite{GiridharKumar}, the problem of worst-case block function computation with zero error was formulated. 
The authors identify two classes of symmetric functions namely \textit{type-sensitive} functions exemplified by Mean, Median and Mode, and \textit{type-threshold} functions, exemplified by Maximum and Minimum. The maximum rates for computation of type-sensitive and type-threshold functions in random planar networks are shown to be $\Theta(\frac{1}{\log n})$ and $\Theta(\frac{1}{\log \log n})$ respectively, where $n$ is the number of nodes. 
For the case of a designated collector node interested in computing the function, one can derive a per-link optimal strategy for block function computation in tree networks \cite{KowshikKumar}.

In contrast, in this paper, we require that every node must compute the function. The latter approach naturally allows the use of tools from communication complexity \cite{KushiNisan}. 
The communication complexity of Boolean functions has been studied in \cite{Wegener}. Further, one can consider the \textit{direct-sum problem} \cite{KarchmerRazWigderson} where several instances of the problem are considered together to obtain savings. This block computation approach is used to compute the exact complexity of the Boolean AND function in \cite{AhlswedeCai}. This result was considerably generalized in \cite{KowshikKumar_ITW} to derive optimal strategies for computing symmetric Boolean functions in broadcast networks.  The average and randomized complexity of Boolean functions are studied in \cite{OrlitskyElgamal}.

While we focus on worst-case computation in this paper, we could suppose that the measurements are drawn from some joint probability distribution. 
The problem of source coding with side information has been studied for the vanishing error case in \cite{WynerZiv}, and for the zero error case  in \cite{AlonOrlitsky}. The problem of source coding for function computation with side information has been studied in \cite{OrlitskyRoche}. However, a tractable information theoretic formulation of the problem of function computation has proved elusive. 
%

In this paper, we allow for all interactive strategies where each node's transmission is allowed to depend on all previous transmissions and the node's measurements. Thus, all network coding strategies \cite{AhlswedeYeung} are subsumed in this class. The rate region for multi-round interactive function computation has been characterized for two nodes \cite{MaIshwar}, and for collocated networks \cite{MaGuptaIshwar}. 

\section{The two node problem}\label{sec_two_node}
Consider two nodes $1$ and $2$ with variables $X_1 \in \{0, 1, \ldots, m_1\}$ and $X_2 \in \{0, 1, \ldots, m_2\}$. Both nodes wish to compute a function $f(X_1, X_2)$ which only depends on the value of $X_1 + X_2$. To put this in context, one can suppose there are $m_1$ Boolean variables collocated at node $1$ and $m_2$ Boolean variables at node $2$, and both nodes wish to compute a symmetric Boolean function of the $n := m_1 + m_2$ variables. We pose the problem in a block computation setting, where each node $i$ has a block of $B$ independent measurements, denoted by $X_i^B$. We consider the class of all interactive strategies, where nodes $1$ and $2$ transmit messages alternately with the value of each subsequent message being allowed to depend on all previous transmissions, and the block of measurements available at the transmitting node. We define a round to include one transmission by each node. A strategy is said to achieve correct block computation if for \textit{every} choice of input $(X_1^B, X_2^B)$, each node $i$ can correctly decode the value of the function block $f^B(X_1, X_2)$ using the sequence of transmissions $b_1, b_2, \ldots$ and its own measurement block $X_i^B$. This is the direct-sum problem in communication complexity.  

Let $\mathcal{S}_B$ be the set of strategies for block length $B$, which achieve zero-error block computation, and let $C(f, S_B, B)$ be the worst-case total number of bits exchanged under strategy $S_B \in \mathcal{S}_B$. The worst-case per-instance complexity of computing a function $f(X_1, X_2)$ is defined as
\begin{displaymath}
C(f) := \lim_{B \rightarrow \infty}\min_{S_B \in \mathcal{S}_B} \frac{C(f, S_B, B)}{B}.
\end{displaymath} 
\subsection{Complexity of sum-threshold functions}
In this paper, we are only interested in functions $f(X_1, X_2)$ which only depend on $X_1 + X_2$. Let us suppose without loss of generality that $m_1 \leq m_2$. We define an interesting class of $\{0,1\}$-valued functions called sum-threshold functions.
\begin{definition}[sum-threshold functions]
A sum-threshold function $\Pi_{\theta}(X_1, X_2)$ with threshold $\theta$ is defined as follows:
\begin{displaymath}
\Pi_{\theta}(X_1, X_2) = \left\{\begin{array}{l}1 \quad \textrm{if } X_1 + X_2 \geq \theta , \\ 0 \quad \textrm{otherwise.}\end{array} \right.
\end{displaymath}
\end{definition}
For the special case where $m_1 = 1, m_2 = 1$ and $\theta = 2$, we recover the Boolean AND function, which was studied in \cite{AhlswedeCai}. Throughout this paper, we will use tools introduced in \cite{AhlswedeCai}. 
\begin{theorem}\label{thm_sum_threshold}
Given any strategy $S_B$ for block computation of the function $\Pi_{\theta}(X_1, X_2)$,

\vspace{-0.1in}\begin{small}
\begin{displaymath}
C(\Pi_{\theta}(X_1, X_2), S_B, B) \geq B\log_{2}\{\min (2\theta +1, 2m_1 + 2, 2(n-\theta + 1) + 1)\}.
\end{displaymath}
\end{small}
Further, there exist single-round strategies $S_B^*$ and $S_B^{**}$, starting with nodes $1$ and $2$ respectively, which satisfy

\vspace{-0.1in}\begin{footnotesize}
\begin{displaymath}
C(\Pi_{\theta}(X_1, X_2), S_B^*, B) \leq \lceil B\log_{2}\{\min (2\theta +1, 2m_1 + 2, 2(n-\theta + 1) + 1)\} \rceil.
\end{displaymath}
\begin{displaymath}
C(\Pi_{\theta}(X_1, X_2), S_B^{**}, B) \leq \lceil B\log_{2}\{\min (2\theta +1, 2m_1 + 2, 2(n-\theta + 1) + 1)\} \rceil.
\end{displaymath}
\end{footnotesize}
\begin{normalsize}
\end{normalsize}
Thus, the complexity of computing $\Pi_{\theta}(X_1, X_2)$ is given by $C(\Pi_{\theta}(X_1, X_2))=\log_{2}\{\min (2\theta +1, 2m_1 + 2, 2(n-\theta + 1) + 1)\}$.
\end{theorem}
\textbf{Proof of achievability:} We consider three cases:\\
\textbf{(a)} Suppose $\theta \leq m_1 \leq m_2$. We specify a strategy $S_B^*$ in which node $1$ transmits first. We begin by observing that inputs $X_1 = \theta, X_1 = (\theta + 1) \ldots, X_1 = m_1$ need not be \textit{separated}, since for each of these values of $X_1$, $\Pi_{\theta}(X_1, X_2) = 1$ for all values of $X_2$. 
Thus node $1$ has an effective alphabet of $\{0, 1, \ldots, \theta\}$. Suppose node $1$ transmits using a prefix-free codeword of length $l(X_1^B)$. At the end of this transmission, node $2$ only needs to indicate one bit for the instances of the block where $X_1 = 0, 1, \ldots, (\theta -1)$. Thus the worst-case total number of bits is 
\begin{displaymath}
L := \max_{X_1^B} (l(X_1^B) + w^{0}(X_1^B) + w^{1}(X_1^B) + \ldots + w^{\theta -1}(X_1^B)),
\end{displaymath}
where $w^{j}(X_1^B)$ is the number of instances in the block where $X_1 = j$. We are interested in finding the codebook which will result in the minimum worst-case number of bits. 
From the Kraft inequality for prefix-free codes we have
\begin{equation}
\sum_{X_1^B \in \{0, 1, \ldots, \theta\}^B}2^{-L + w^{0}(X_1^B) + w^{1}(X_1^B) + \ldots + w^{\theta -1}(X_1^B))} \leq 1. \nonumber
\end{equation}  
Consider a codebook with $l(X_1^B) = \lceil B\log_{2}(2\theta +1)\rceil - w(x_1^B)$. This satisfies the Kraft inequality since 
\begin{displaymath}
\sum_{X_1^B \in \{0, 1, \ldots, \theta\}^B}2^{w^{0}(X_1^B) + w^{1}(X_1^B) + \ldots + w^{\theta -1}(X_1^B))}.1^{w^{\theta}(X_1^B)} = (2\theta + 1)^B.
\end{displaymath}
Hence there exists a prefix-free code for which the worst-case total number of bits exchanged is $\lceil B\log_{2}(2\theta +1) \rceil$. Since $\theta \leq m_1 \leq m_2$, we have

\vspace{-0.1in}\begin{footnotesize}
\begin{displaymath}
C(\Pi_{\theta}(X_1, X_2), S_B^*, B) \leq \lceil B\log_{2}\{\min (2\theta +1, 2m_1 + 2, 2(n-\theta + 1) + 1)\}\rceil.
\end{displaymath}
\end{footnotesize}
\begin{normalsize}
\end{normalsize}
The strategy $S_B^{**}$ starting at node $2$ can be similarly derived. Node $2$ now has an effective alphabet of $\{0, 1, \ldots, \theta\}$, and we have $C(\Pi_{\theta}(X_1, X_2), S_B^{**}, B) \leq \lceil B\log_{2} (2\theta +1) \rceil$.\\
\textbf{(b)} Suppose $m_1 \leq m_2 < \theta$. Consider a strategy $S_B^*$ in which node $1$ transmits first. The inputs $X_1 = 0, X_1 = 1, \ldots, X_1 = \theta - m_2 -1$ need not be \textit{separated} since for each of these values of $X_1$, $\Pi_{\theta}(X_1, X_2) = 0$ for all values of $X_2$. Thus node $1$ has an effective alphabet of $\{\theta - m_2 - 1, \theta - m_2, \ldots, m_1\}$. Upon hearing node $1$'s transmission, node $2$ only needs to indicate one bit for the instances of the block where $X_1 = \theta - m_2, \ldots, m_1$. Consider a codebook with $l(X_1^B) = \lceil B\log_{2}(2(m_1 + m_2 - \theta +1) + 1)\rceil - w^{\theta - m_2}(X_1^B) - \ldots - w^{m_1}(X_1^B)$. This satisfies the Kraft inequality and we have $L = \lceil B \log_{2}(2(n- \theta + 1) + 1)\rceil$. Since $m_1 \leq m_2 < \theta$, we have that

\vspace{-0.1in}\begin{footnotesize}
\begin{displaymath}
C(\Pi_{\theta}(X_1, X_2), S_B^*, B) \leq \lceil B\log_{2}\{\min (2\theta +1, 2m_1 + 2, 2(n-\theta + 1) + 1)\}\rceil .
\end{displaymath}
\end{footnotesize}
The strategy $S_B^{**}$ starting at node $2$ can be analogously derived. \\
\textbf{(c)} Suppose $m_1 < \theta \leq m_2$. For the case where node $1$ transmits first, we construct a trivial strategy $S_B^*$ where node $1$ uses a codeword of length $\lceil B\log_{2}(m_1 + 1)\rceil$ bits and node $2$ replies with a string of $B$ bits indicating the function block. Thus we have $C(\Pi_{\theta}(X_1, X_2), S_B^*, B) \leq \lceil B\log_{2}(2m_1 + 2)\rceil$. 

Now consider a strategy $S_B^{**}$ where node $2$ transmits first. Observe that the inputs $X_2 = 0, X_2 = 1, \ldots, X_2 = \theta - m_1 -1$ need not be separated since for each of these values of $X_2$, $\Pi_{\theta}(X_1, X_2) = 0$ for all values of $X_2$. Further, the inputs $X_2 = \theta , X_2 = \theta + 1, \ldots, X_2 = m_2$ need not be \textit{separated}. Thus node $1$ has an effective alphabet of $\{\theta - m_1 - 1, \theta - m_1, \ldots, \theta\}$. Upon hearing node $2$'s transmission, node $1$ only needs to indicate one bit for the instances of the block where $X_2 = \theta - m_1, \ldots, \theta - 1$. Consider a codebook with $l(X_2^B) = \lceil B\log_{2}(2m_1 + 2)\rceil - w^{\theta - m_1}(X_1^B) - \ldots - w^{\theta -1}(X_1^B)$. This satisfies the Kraft inequality and we have $L = \lceil B\log_{2}(2(n - \theta +1) + 1) \rceil$. Since $m_1 < \theta \leq m_2$, we have that

\vspace{-0.1in}\begin{footnotesize}
\begin{displaymath}
C(\Pi_{\theta}(X_1, X_2), S_B^{**}, B) \leq \lceil B\log_{2}\{\min (2\theta +1, 2m_1 + 2, 2(n-\theta + 1) + 1)\}\rceil .
\end{displaymath}
\end{footnotesize}

\vspace{-0.1in}The lower bound is shown by constructing a \textit{fooling set} \cite{KushiNisan}.
\begin{definition}[Fooling Set]
A set $E \subseteq \mathcal{X} \times \mathcal{Y}$ is said to be a fooling set, if for any two distinct elements $(x_1, y_1), (x_2, y_2)$ in $E$, we have either 
\begin{itemize}
\item$f(x_1, y_1) \neq f(x_2, y_2)$, or
\item$f(x_1, y_1) = f(x_2, y_2)$, but either $f(x_1,y_2) \neq f(x_1, y_1)$ or $f(x_2,y_1) \neq f(x_1,y_1)$.
\end{itemize}
\end{definition}
Given a fooling set $E$ for a function $f(X_1, X_2)$, we have $C(f(X_1, X_2)) \geq \log_{2}|E|$.\\
\textbf{Proof of Lower Bound:} Define the measurement matrix $M$ to be the matrix obtained by stacking the row $X_1^B$  over the row $X_2^B$. Let $E$ denote the set of all measurement matrices which are made up only of the column vectors from the set 

\vspace{-0.1in}\begin{small}
\begin{displaymath}
Z = \left\{\left[\begin{array}{c} z_1 \\ z_2 \end{array}\right]: 0 \leq z_1 \leq m_1, 0 \leq z_2 \leq m_2, (\theta -1) \leq z_1 + z_2 \leq \theta \right\}.
\end{displaymath}
\end{small}

We claim that $E$ is the appropriate fooling set. Consider two distinct measurement matrices $M_1, M_2 \in E$. Let $f^B(M_1)$ and $f^B(M_2)$ be the block function values obtained from these two matrices. If $f^B(M_1) \neq f^B(M_2)$, we are done. Let us suppose $f^B(M_1) = f^B(M_2)$, and note that since $M_1 \neq M_2$, there must exist one column where $M_1$ and $M_2$ differ. Suppose $M_1$ has $\left[\small{\begin{array}{c} z_{1a} \\ z_{2a} \end{array}}\right]$ while $M_2$ has $\left[\small{\begin{array}{c} z_{1b} \\ z_{2b} \end{array}}\right]$, where $z_{1a} + z_{2a} = z_{1b} + z_{2b}$. Assume without loss of generality that $z_{1a} < z_{1b}$ and $z_{2a} > z_{2b}$. 
\begin{itemize}
\item If $z_{1a} + z_{2a} = z_{1b} + z_{2b} = \theta -1$, then the \textit{diagonal} element $f(z_{1b}, z_{2a}) = 1$ since $z_{1b} + z_{2a} \geq \theta$. Thus, if we replace the first row of $M_1$ with the first row of $M_2$, the resulting measurement matrix, say $M^{*}$, is such that $f(M^{*}) \neq f(M_1)$.
\item If $z_{1a} + z_{2a} = z_{1b} + z_{2b} = \theta$, then the \textit{diagonal} element $f(z_{1a}, z_{2b}) = 0$ since $z_{1b} + z_{2a} < \theta$. Thus, if we replace the second row of $M_1$ with the second row of $M_2$, the resulting matrix $M^{*}$ is such that $f(M^{*}) \neq f(M_1)$.
\end{itemize}
Thus, the set $E$ is a valid fooling set with cardinality $|Z|^B$. For any strategy $S_B$, we have $C(f, S_B, B) \geq B\log_{2}|Z|$. The cardinality of $Z$ can be modeled as the sum of the coefficients of $Y^{\theta}$ and $Y^{\theta -1 }$ in a carefully constructed polynomial:
\begin{displaymath}
|Z| = \left[Y^{\theta}\right] + \left[Y^{\theta -1 }\right](1 + Y + \ldots + Y^{m_1})(1 + Y + \ldots + Y^{m_2})
\end{displaymath}
This is solved using the binomial expansion for $\frac{1}{(1-Y)^k}$ \cite{West}.
\begin{itemize}
\item[(a)] Suppose $\theta \leq m_1 \leq m_2$. Then $|Z| = \theta + \theta + 1$.
\item[(b)] Suppose $m_1 \leq \theta \leq m_2$. Then $|Z| = 2m_1 + 2$.
\item[(c)] Suppose $m_1 \leq m_2 \leq \theta$. Then $|Z| = 2(n- \theta + 1) +1$.
\end{itemize}
This completes the proof of Theorem \ref{thm_sum_threshold}. $\Box$ 
\subsection{Complexity of sum-interval functions}
\begin{definition}[sum-interval functions]
A sum-interval function $\Pi_{[a,b]}(X_1, X_2)$ on the interval $[a,b]$ is defined as follows:
\begin{displaymath}
\Pi_{[a,b]}(X_1, X_2) := \left\{\begin{array}{l}1 \quad \textrm{if } a \leq X_1 + X_2 \leq b, \\ 0 \quad \textrm{otherwise.} \end{array}\right.
\end{displaymath}
\end{definition}
\begin{theorem}\label{thm_sum_interval}
Given any strategy $S_B$ for block computation of $\Pi_{[a,b]}(X_1, X_2)$ where $b \leq n/2$,
\begin{displaymath}
C(\Pi_{[a,b]}(X_1, X_2), S_B, B) \geq B\log_{2}\{\min (2b-a+3, m_1 + 1)\}.
\end{displaymath}
Further, there exists a single-round strategy $S_B^*$ which satisfies
\begin{displaymath}
C(\Pi_{[a,b]}(X_1, X_2), S_B^*, B) \leq \lceil B\log_{2}\{\min (2(b+1)+1, 2m_1 + 2)\} \rceil .
\end{displaymath}
Thus, we have obtained the complexity of computing $\Pi_{\theta}(X_1, X_2)$ to within one bit.
\end{theorem}
\subsection{A general strategy for achievability}
The strategy for achievability used in Theorems \ref{thm_sum_threshold} and \ref{thm_sum_interval} suggests an achievable scheme for any general function $f(X_1, X_2)$ of variables $X_1 \in \mathcal{X}_1$ and $X_2 \in \mathcal{X}_2$ which depends only on the value of $X_1 + X_2$. This is done in two stages.\\
\textbf{Separation:} Two inputs $x_{1a}$ and $x_{1b}$ need not be \textit{separated} if $f(x_{1a}, x_2) = f(x_{1b}, x_2)$ for all values $x_2$. By checking this condition for each pair $(x_{1a}, x_{1b})$, we can arrive at a partition of $\{0, 1 \ldots, m_1\}$ into equivalence classes, which can be considered a reduced alphabet, say $A := \{a_1, \ldots, a_l\}$. \\
\textbf{Coding:} Let $A_{0}$ denote the subset of the alphabet $A$ for which the function evaluates only to $0$, irrespective of the value of $X_2$, and let $A_{1}$ denote the subset of $A$ which always evaluates to $1$. Clearly, from the equivalence class structure, we have $|A_{0}| \leq 1$ and $|A_{1}| \leq 1$. Using the Kraft inequality as in Theorems \ref{thm_sum_threshold} and \ref{thm_sum_interval}, we obtain a scheme $S_B^*$ with complexity $\log_{2}(2l - |A_0| - |A_1|)$.
\section{Computing symmetric Boolean functions on tree networks}\label{sec_trees}
Consider a tree graph $T = (V, E)$, with node set $V = \{0, 1, \ldots, n\}$ and edge set $E$. Each node $i$ has a Boolean variable $X_i \in \{0,1\}$, and every node wants to compute a given symmetric Boolean function $f(X_1, X_2, \ldots, X_n)$. Again, we allow for block computation and consider all strategies where nodes can transmit in any sequence with possible repititions, subject to:
\begin{itemize}
\item On any edge $e = (i,j)$, either node $i$ transmits or node $j$ transmits, or neither, and this is determined from the previous transmissions.
\item Node $i$'s transmission can depend on the previous transmissions and the measurement block $X_i^B$.
\end{itemize} 
For sum-threshold functions, we have a computation and communication strategy that is optimal for each link. 
\begin{theorem}\label{thm_tree_threshold}
Consider a tree network where we want to compute the function $\Pi_{\theta}(X_1, \ldots, X_n)$. Let us focus on a single edge $e \equiv (i,j)$ whose removal disconnects the graph into components $A_{e}$ and $V\setminus A_{e}$, with $|A_{e}| \leq |V \setminus A_{e}|$. For any strategy $S_B \in \mathcal{S}_B$, the number of bits exchanged along edge $e \equiv (i,j)$, denoted by $C_{e}(\Pi_{\theta}(X_1, \ldots, X_n), S_B, B)$, is lower bounded by

\vspace{-0.1in}\begin{footnotesize}
\begin{equation}
C_{e}(\Pi_{\theta}(X_1, \ldots, X_n), S_B, B) \\
\geq B\log_{2}\{\min (2\theta +1, 2|A_{e}| + 2, 2(n-\theta + 1) + 1)\}. \nonumber
\end{equation}
\end{footnotesize}
\normalsize{Further, there exists a strategy $S_B^*$ such that for any edge $e$,} 
\begin{footnotesize}
\begin{equation}
C_{e}(\Pi_{\theta}(X_1, \ldots, X_n), S_B^*, B) \\
\leq \lceil B\log_{2}\{\min (2\theta +1, 2|A_{e}| + 2, 2(n-\theta + 1) + 1)\}\rceil. \nonumber
\end{equation}
\end{footnotesize}
\normalsize{The complexity of computing $\Pi_{\theta}(X_1, \ldots, X_n)$ is given by}
\begin{small}
\begin{displaymath}
C_{e}(\Pi_{\theta}(X_1, \ldots, X_n)) = \log_{2}\{\min (2\theta +1, 2|A_{e}| + 2, 2(n-\theta + 1) + 1)\}.
\end{displaymath}
\end{small}
\end{theorem}
\textbf{Proof:} Given a tree network $T$, every edge $e$ is a cut edge. Consider an edge $e$ whose removal creates components $A_{e}$ and $V \setminus A_{e}$, with $|A_{e}| \leq  |V \setminus A_{e}|$. Now let us aggregate the nodes in $A_{e}$ and also those in $V \setminus A_{e}$, and view this as a problem with two nodes connected by edge $e$. Clearly the complexity of computing the function $\Pi_{\theta}(X_{A_{e}}, X_{V \setminus A_{e}})$ is a lower bound on the worst-case total number of bits that must be exchanged on edge $e$ under any strategy $S_B$. Hence we obtain

\vspace{-0.1in}\begin{footnotesize}
\begin{equation}
C_{e}(\Pi_{\theta}(X_1, \ldots, X_n), S_B, B) \geq B\log_{2}\{\min (2\theta +1, 2|A_{e}| + 2, 2(n-\theta + 1) + 1)\}. \nonumber
\end{equation}
\end{footnotesize}
The achievable strategy $S_B^*$ is derived from the achievable strategy for the two node case in Theorem \ref{thm_sum_threshold}. While the transmissions back and forth along any edge will be exactly the same, we need to orchestrate these transmissions so that conditions of causality are maintained. Pick any node, say $r$, to be the root. This induces a partial order on the tree network. We start with each leaf in the network transmitting its codeword to the parent. Once a parent node obtains a codeword from each of its children, it has sufficient knowledge to disambiguate the letters of the effective alphabet of the subtree, and subsequently it transmits a codeword to its parent. Thus codewords are transmitted from child nodes to parent nodes until the root is reached. The root can then compute the value of the function and now sends the appropriate replies to its children. The children then compute the function and send appropriate replies, and so on. This sequential strategy  depends critically on the fact that, in the two node problem, we derived optimal strategies starting from either node. For any edge $e$, the worst-case total number of bits exchanged is given by

\vspace{-0.1in}\begin{footnotesize}
\begin{equation}
C_{e}(\Pi_{\theta}(X_1, \ldots, X_n), S_B^*, B) \leq \lceil B\log_{2}\{\min (2\theta +1, 2|A_{e}| + 2, 2(n-\theta + 1) + 1)\}\rceil . \Box \nonumber
\end{equation}
\end{footnotesize}
One can similarly derive an approximately optimal strategy for sum-interval functions, which we state here without proof.
\begin{theorem} \label{thm_tree_interval}
Consider a tree network where we want to compute the function $\Pi_{[a,b]}(X_1, \ldots, X_n)$, with $b \leq \frac{n}{2}$. Let us focus on a single edge $e \equiv (i,j)$ whose removal disconnects the graph into components $A_{e}$ and $V\setminus A_{e}$, with $|A_{e}| \leq |V \setminus A_{e}|$. For any strategy $S_B \in \mathcal{S}_B$, the number of bits exchanged along edge $e \equiv (i,j)$, denoted by $C_{e}(f, S_B, B)$ is lower bounded by 
\begin{displaymath}
C_{e}(\Pi_{[a,b]}(X_1, \ldots, X_n), S_B, B) \geq B\log_{2}\{\min (2b-a+3, |A_e| + 1)\}. 
\end{displaymath}
Further there exists a strategy $S_B^*$ such that for any edge $e$,
\begin{small}
\begin{displaymath}
C_{e}(\Pi_{[a,b]}(X_1, \ldots, X_n), S_B^*, B) \leq \lceil B\log_{2}\{\min (2(b+1) +1, 2|A_e| + 2)\}\rceil.
\end{displaymath}
\end{small}
\end{theorem}
\subsection{Extension to non-binary alphabets}
The extension to the case where each node draws measurements from a non-binary alphabet is immediate. 
Consider a tree network with $n$ nodes where node $i$ has a measurement $X_i \in \{0, 1, \ldots, l_i -1\}$. Suppose all nodes want to compute a given function which only depends on the value of $X_1 + X_2 + \ldots + X_n$. We can define sum-threshold functions in analogous fashion and derive an optimal strategy for computation. 
\begin{theorem}
Consider a tree network where we want to compute a sum-threshold function, $\Pi_{\theta}(X_1, \ldots, X_n)$, of non-binary measurements. Let us focus on a single edge $e$ whose removal disconnects the graph into components $A_{e}$ and $V\setminus A_{e}$. Let us define $l_{A_e} := \sum_{i \in A_e} l_i$. Then the complexity of computing $\Pi_{\theta}(X_1, \ldots, X_n)$ is given by 

\vspace{-0.1in}\begin{footnotesize}
\begin{equation}
C_{e}(\Pi_{\theta}(X_1, \ldots, X_n)) \\
= \log_{2}\{\min (2\theta +1, 2\min(l_{A_e}, l_{V \setminus A_e}) + 2, 2(l_{V}-\theta + 1) + 1)\}. \nonumber
\end{equation}
\end{footnotesize}
\end{theorem}
Theorem \ref{thm_tree_interval} also extends to the case of non-binary alphabets.
\section{Computing sum-threshold functions in general graphs}\label{sec_gen_graphs}
We now consider the computation of sum-threshold functions in general graphs where the alphabet is not restricted to be binary. A cut is defined to be a set of edges $F \subseteq E$ which disconnect the network into two components $A_F$ and $V \setminus A_F$. 
\begin{lemma}[Cut-set bound]
Consider a general network $G = (V, E)$, where node $i$ has measurement $X_i \in \{0, 1, \ldots, l_i - 1\}$ and all nodes want to compute the function $\Pi_{\theta}(X_1, \ldots, X_n)$. Given a cut $F$ which separates $A_F$ from $V \setminus A_F$, the cut-set lower bound specifies that: For any strategy $S_B$, the number of bits exchanged on the edges in $F$ is lower bounded by

\vspace{-0.1in}\begin{footnotesize}
\begin{equation}
C_{F}(\Pi_{\theta}(X_1, \ldots, X_n), S_B, B) \geq B \log_{2}(\min \{2\theta +1, 2m_F + 2, 2(l_{V} - \theta + 1) + 1)\}. \nonumber
\end{equation}
\end{footnotesize}
where $l_{A_F} = \sum_{i \in A_F} l_i$ and $m_F = \min(l_{A_F}, l_{V \setminus A_F})$.
\end{lemma}

A natural achievable strategy is to pick a spanning subtree of edges and use the optimal strategy on this subtree. The convex hull of the rate vectors of the subtree aggregation schemes, is an achievable region. We wish to compare this with the cut-set region. To simplify matters, consider a complete graph $G$ where each node $i$ has a measurement $X_i \in \{0, \ldots, l-1\}$. Let $R_{ach}$ be the maximum symmetric ratepoint achievable by aggregating along trees, and $R_{cut}$ be the minimum symmetric ratepoint that satisfies the cut-set constraints. 
\begin{theorem} For the computation of sum-threshold functions on complete graphs, $R_{ach} \leq 2(1 - \frac{1}{n}))R_{cut}$. In fact, this approximation ratio is tight.
\end{theorem}
\textbf{Proof:} Let us assume without loss of generality that $\theta \leq \frac{n.l}{2}$. Consider all cuts of the type $(\{i\}, V \setminus \{i\})$. This yields 
\begin{displaymath}
R_{cut} \geq \max_{i \in V}\left(\frac{\min(\log_{2}(2\theta + 1), \log_{2}(2l_i+2))}{n-1}\right).
\end{displaymath}
Now consider the achievable scheme which employs each of the $n$ star graphs for equal sized sub-blocks of measurements. 
The rate on edge $(i,j)$ is given by

\vspace{-0.1in}\begin{footnotesize}
\begin{displaymath}
\frac{1}{n}\left(\min(\log_{2}(2\theta + 1), \log_{2}(2l_i+2)) + \min(\log_{2}(2\theta + 1), \log_{2}(2l_j+2))\right)
\end{displaymath}
\end{footnotesize}
\normalsize{Hence we have}
\begin{small}
\begin{equation}
R_{ach} \leq \frac{2}{n}(\min(\log_{2}(2\theta + 1), \max_{i \in V}\{\log_{2}(2l_i+2)\})) \leq 2\left(1 - \frac{1}{n}\right)R_{cut}. \nonumber
\end{equation}
\end{small}
\section{Concluding Remarks}
In this paper, we have addressed the computation of symmetric Boolean functions in tree networks, where all nodes want to compute the function. Toward this objective, we derived lower bounds on the number of bits that must be exchanged on each edge, using communication complexity theory. Further, for each edge, we devise an achievable scheme for block computation that involves separation followed by prefix-free coding. We then sequence the transmissions so that information flows up the tree to a root node and then back down to the leaves. For the case of sum-threshold functions, our resulting achievable scheme is optimal. 

The approach presented also provides lower and upper bounds for the complexity of other functions like sum-interval functions. Our framework can be generalized to handle functions of integer measurements which only depend on the sum of the measurements. The extension to general graphs is very interesting and appears significantly harder. However, a cut-set lower bound can be immediately derived, and in some special cases we can show that subtree aggregation schemes provide a 2-OPT solution.
\bibliographystyle{unsrt}
\bibliography{isit_biblio}
\end{document}